\documentclass{llncs}

\usepackage{makeidx}

\usepackage{amsmath}
\usepackage{amssymb}
\usepackage{stmaryrd}
\usepackage{pifont}
\usepackage[all]{xy}

\newcommand{\keywords}[1]{\par\addvspace\baselineskip
\noindent\keywordname\enspace\ignorespaces#1}

\begin{document}

\mainmatter              % start of the contributions

%----------------------------------------------------------------------------------------------
% Document description
%----------------------------------------------------------------------------------------------

\title{Why Would You Trust \emph{B}?}

% abbreviated title (for running head) also used for the TOC unless \toctitle is used
\titlerunning{\emph{B} in \emph{Coq}}

\author{\'Eric Jaeger\inst{1}\inst{2} \and Catherine Dubois\inst{3}}

% abbreviated author list (for running head)
\authorrunning{Jaeger, Dubois}

% modified list of authors for the TOC (add the affiliations)
\tocauthor{
Author1 (Entity1),
Author2 (Entity2)
}
\institute{
LIP6, Universit\'e Paris 6,
4 place Jussieu, 75252 Paris Cedex 05, France
\and
LTI, Direction centrale de la s\'ecurit\'e des syst\`emes d'information,
51 boulevard de La Tour-Maubourg, 75700 Paris 07 SP, France
%\email{eric.jaeger@sgdn.pm.gouv.fr}
\and
CEDRIC, \'Ecole nationale sup\'erieure d'informatique pour l'industrie et l'entreprise,
18 all\'ee Jean Rostand, 91025 Evry Cedex, France
%\email{dubois@iie.fr}
}

% typeset the title of the contribution
\maketitle

%----------------------------------------------------------------------------------------------
% Abstract
%----------------------------------------------------------------------------------------------

\begin{abstract}
The use of formal methods provides confidence in the correctness of developments. Yet one may 
argue about the actual level of confidence obtained when the method itself -- or its 
implementation -- is not formally checked. We address this question for the \emph{B}, a widely 
used formal method that allows for the derivation of correct programs from specifications. 
Through a deep embedding of the \emph{B} logic in \emph{Coq}, we check the \emph{B} theory but 
also implement \emph{B} tools. Both aspects are illustrated by the description of a proved 
prover for the \emph{B} logic.
\end{abstract}

%----------------------------------------------------------------------------------------------
% Macros
%----------------------------------------------------------------------------------------------

\newcommand{\NAT}{\ensuremath{\mathbb{N}}}

\newcommand{\choice}{\ensuremath{\talloblong}}
\newcommand{\guard}{\ensuremath{\Longrightarrow}}

\newcommand{\rulename}[1]{\ensuremath{\:{\scriptstyle{[{#1}]}}}}

\newcommand{\Bidx}{\ensuremath{\mathbb{I}}}
\newcommand{\dbzero}{\ensuremath{\text{0}}}
\newcommand{\dbone}{\ensuremath{\text{1}}}
\newcommand{\dbtwo}{\ensuremath{\text{2}}}
\newcommand{\dbthree}{\ensuremath{\text{3}}}
\newcommand{\dbfour}{\ensuremath{\text{4}}}
\newcommand{\dbfive}{\ensuremath{\text{5}}}

\newcommand{\Bnamp}{\ensuremath{\mathbb{K}}}
\newcommand{\Bnamo}{\ensuremath{\mathbb{J}}}

\newcommand{\Bbol}{\ensuremath{\mathbb{B}}}

\newcommand{\Bprd}{\ensuremath{\mathbb{P}}}
\newcommand{\Bexp}{\ensuremath{\mathbb{E}}}
\newcommand{\Btrm}{\ensuremath{\mathbb{T}}}
\newcommand{\pand}{\ensuremath{\dot{\land}}}
\newcommand{\pimp}{\ensuremath{\dot{\Rightarrow}}}
\newcommand{\pnot}{\ensuremath{\dot{\lnot}}}
\newcommand{\pfor}{\ensuremath{\dot{\forall}\:}}
\newcommand{\pequ}{\ensuremath{\dot{=}}}
\newcommand{\pins}{\ensuremath{\dot{\in}}}
\newcommand{\ppvr}[1]{\ensuremath{\dot{\pi}_{#1}}}
\newcommand{\evar}[1]{\ensuremath{\dot{\chi}_{#1}}}
\newcommand{\ecpl}{\ensuremath{\dot{\mapsto}}}
\newcommand{\echs}{\ensuremath{\dot{\downarrow} \;\!\!}}
\newcommand{\epro}{\ensuremath{\dot{\times}}}
\newcommand{\epow}{\ensuremath{\dot{\uparrow} \;\!\!}}
\newcommand{\ecmp}[2]{\ensuremath{\{ {#1}\dot{|}{#2} \} }}
\newcommand{\ebig}{\ensuremath{\dot{\Omega}}}
\newcommand{\ebie}[1]{\ensuremath{\dot{\omega}_{#1}}}
\newcommand{\piff}{\ensuremath{\dot{\Leftrightarrow}}}
\newcommand{\por}{\ensuremath{\dot{\lor}}}
\newcommand{\pexs}{\ensuremath{\dot{\exists}}}

\newcommand{\Bnotfree}{\ensuremath{\dot{\smallsetminus}}}

\newcommand{\DBinter}[1]{\left\llbracket {#1} \right\rrbracket}

\newcommand{\fresh}{\ensuremath{\mathcal{F}}}
\newcommand{\depth}{\ensuremath{\mathcal{D}}}

\newcommand{\Bcmd}[1]{\ensuremath{\mathbf{#1}}}

\newcommand{\Bbdfor}[2]{\ensuremath{\!\uparrow_{\!\forall\!}\!({#1}\!\cdot\!{#2})}}
\newcommand{\Bbdexs}[2]{\ensuremath{\!\uparrow_{\!\exists\!}\!({#1}\!\cdot\!{#2})}}
\newcommand{\Bbdcmp}[3]{\ensuremath{\!\uparrow_{\!\{\!\}}\!\!({#1}\!:\!{#2}\!\cdot\!{#3})}}
\newcommand{\Binfor}[2]{\ensuremath{\!\downarrow_{\forall\!}\!({#2}\!\gets\!{#1})}}
\newcommand{\Binexs}[2]{\ensuremath{\!\downarrow_{\exists\!}\!({#2}\!\gets\!{#1})}}
\newcommand{\Bincmp}[2]{\ensuremath{\!\downarrow_{\{\!\}}\!\!({#2}\!\gets\!{#1})}}
\newcommand{\Baffec}[3]{\ensuremath{\langle{#1}\!:=\!{#2}\rangle{#3}}}

\newcommand{\Bmaf}{\ensuremath{\mathbb{M}}}
\newcommand{\Bmaffec}[2]{\ensuremath{\langle\!\langle{#1}|{#2}\rangle\!\rangle}}

\newcommand{\Baffprd}[3]{\ensuremath{\langle{#1}\!:\equiv\!{#2}\rangle{#3}}}
\newcommand{\Bgraftprd}[3]{\ensuremath{\langle{#1}\!\lhd\!{#2}\rangle{#3}}}

\newcommand{\Binf}{\ensuremath{\:\dot{\vdash}\:}}

\spnewtheorem*{notation}{Notation}{\bfseries}{\itshape}
\spnewtheorem*{defn}{Definition}{\bfseries}{\itshape}

%----------------------------------------------------------------------------------------------
% Main body
%----------------------------------------------------------------------------------------------

A clear benefit of formal methods is to increase the confidence in the correctness of 
developments. However, one may argue about the actual level of confidence obtained, when the 
method or its implementation are not themselves formally checked. This question is legitimate 
for safety, as one may accidentally derive invalid results. It is even more relevant when 
security is a concern, as any flaw can be deliberately exploited by a malicious developer to 
obfuscate undesirable behaviours of a system while still getting a certification.

\emph{B} \cite{abr:1} is a popular formal method that allows for the derivation of correct 
programs from specifications. Several industrial implementations are available (e.g. 
\emph{AtelierB}, \emph{B Toolkit}), and it is widely used in the industry for projects where 
safety or security is mandatory. So the \emph{B} is a good candidate for addressing our 
concern: when the prover says that a development is right, who says that the prover is right? 
To answer this question, one has to check the theory as well as the prover w.r.t. this theory 
(or, alternatively, to provide a proof checker). Those are the objectives of \emph{BiCoq}, a 
deep embedding of the \emph{B} logic  in \emph{Coq} \cite{coq:1}.

\emph{BiCoq} benefits from the support of \emph{Coq} to study the theory of \emph{B}, and to 
check the validity of standard definitions and results. \emph{BiCoq} also allows us, through 
an implementation strategy, to develop formally checked \emph{B} tools. This strategy is 
illustrated in this paper by the development of a prover engine for the \emph{B} logic, that 
can be extracted and used independently of \emph{Coq}. \emph{Coq} is therefore our notary 
public, witnessing the validity of the results associated to the \emph{B} theory, as well as 
the correctness of tools implementing those results -- ultimately increasing confidence in 
\emph{B} developments. The approach, combining a deep embedding and an implementation 
technique, can be extended to address further elements of the \emph{B}, beyond its logic, or 
to safely enrich it, as illustrated in this paper.

This paper is divided into 9 sections. Sections \ref{sc_introb}, \ref{sc_introcoq} and
\ref{sc_relworks} briefly introduce \emph{B}, \emph{Coq} and the notion of embedding. The 
\emph{B} logic and its formalisation in \emph{Coq} are presented in Sec. \ref{sc_formal}. 
Section \ref{sc_proofs} describes various results proved using \emph{BiCoq}. Section 
\ref{sc_environ} focuses on the implementation strategy, and presents its application to the 
development of a set of extractible proof tactics for a \emph{B} prover. Section 
\ref{sc_extend} discusses further uses of \emph{BiCoq}, and mentions some existing extensions. 
Finally, Sect. \ref{sc_conc} concludes and identifies further activities.

\section{A Short Introduction to \emph{B}}\label{sc_introb}

In a nutshell, the \emph{B} method defines a first-order predicate logic completed with 
elements of set theory, a \emph{Generalised Substitution Language} (\emph{GSL}) and a 
methodology of development. An abstract \emph{B machine} is a module combining a state, 
properties and operations (described as substitutions) to read or alter the state.

The logic is used to express preconditions, invariants, etc. and to conduct proofs. The 
\emph{GSL} allows for definitions of substitutions that can be abstract, declarative and 
non-deterministic (that is, specifications) as well as concrete, imperative and deterministic 
(that is, programs). The following example uses the non-deterministic substitution 
$\Bcmd{ANY}$ (a ``magic'' operator finding a value which satisfies a property) to specify the 
square root of a natural number $n$:
\begin{example}\small$\Bcmd{ANY}\:x\:\Bcmd{WHERE}\:x\!*\!x\leq\!n\!<\!(x\!+\!1)\!*\!(x\!+\!1)\:
\Bcmd{THEN}\:\sqrt(n)\!:=\!x\:\Bcmd{END}$
\end{example}

Regarding the methodology, a machine $M_C$ \emph{refines} an abstract machine $M_A$ if one 
cannot distinguish $M_C$ from $M_A$ by valid operation calls -- this notion being independent 
of the internal representations, as illustrated by the following example of a system returning 
the maximum of a set of stored values:
\begin{example}\small The state of $M_A$ is a (non implementable) set of natural numbers; the 
state of $M_C$ is a natural number. Yet $M_C$, having the expected behaviour, refines $M_A$.
\[
\begin{array}{lll}
\begin{array}{l}
\Bcmd{MACHINE}\:M_A\\
\Bcmd{VARIABLES}\:S\\
\Bcmd{INVARIANT}\:S\!\subseteq\!\NAT\\
\Bcmd{INITIALISATION}\:S\!:=\!\emptyset\\
\Bcmd{OPERATIONS}\\
store(n)\triangleq\\
\quad\Bcmd{PRE}\:n\!\in\!\NAT\:\Bcmd{THEN}\:S\!:=\!S\!\cup\!\{n\}\:\Bcmd{END}\\
m\!\gets\!get\triangleq\\
\quad\Bcmd{PRE}\:S\!\not=\!\emptyset\:\Bcmd{THEN}\:m\!:=\!max(S)\:\Bcmd{END}\\
\Bcmd{END}
\end{array}
& \quad &
\begin{array}{l}
\Bcmd{REFINEMENT}\:M_C\\
\Bcmd{VARIABLES}\:s\\
\Bcmd{INVARIANT}\:s\!=\!max(S\!\cup\!\{0\})\\
\Bcmd{INITIALISATION}\:s\!:=\!0\\
\Bcmd{OPERATIONS}\\
store(n)\triangleq\\
\quad\Bcmd{IF}\:s\!<\!n\:\Bcmd{THEN}\:s\!:=\!n\:\Bcmd{END}\\
m\!\gets\!get\triangleq\\
\quad\Bcmd{BEGIN}\:m\!:=\!s\:\Bcmd{END}\\
\Bcmd{END}
\end{array}
\end{array}
\]
\end{example}
Refinement being transitive, it is possible to go progressively from the specification to the 
implementation. By discharging at each step the \emph{proof obligations} defined by the 
\emph{B} methodology, a program can be proved to be a correct and complete implementation of a 
specification. This methodology, combined with the numerous native notions provided by the set 
theory and the existence of toolkits, make the \emph{B} a popular formal method, widely used 
in the industry.

Note that the \emph{B} logic is not genuinely typed and allows for manipulation of free 
variables. A special mechanism, called \emph{type-checking} (but thereafter referred to as 
\emph{wf-checking}), filters ill-formed (potentially paradoxal) terms; it is only mentioned 
in this paper, deserving a dedicated analysis.

The rest of the paper only deals with the \emph{B} logic (its inference rules).

\section{A Short Introduction to \emph{Coq}}\label{sc_introcoq}

\emph{Coq} is a proof assistant based on a type theory. It offers a higher-order logical 
framework that allows for the construction and verification of proofs, as well as the
development and analysis of functional programs in an \emph{ML}-like language with 
pattern-matching. It is possible in \emph{Coq} to define values and types, including dependent 
types (that is, types that explicitly depend on values); types of sort $\Bcmd{Set}$ represent 
sets of computational values, while types of sort $\Bcmd{Prop}$ represent logical 
propositions. When defining an inductive type (that is, a least fixpoint), associated 
structural induction principles are automatically generated.

For the intent of this paper, it is sufficient to see \emph{Coq} as allowing for the
manipulation of inductive sets of terms. For example, let's consider the standard 
representation of natural numbers:
\begin{example}\small
$\Bcmd{Inductive}\:\NAT\!:\!\Bcmd{Set}\!:=\!0\!:\!\NAT\:|\:S\!:\!\NAT\!\to\!\NAT$
\end{example}
It defines a type $\NAT$ which is the smallest set of terms stable by application of the 
constructors $0$ and $S$. $\NAT$ is exactly made of the terms $0$ and $S^n(0)$ for any finite 
$n$; being well-founded, structural induction on $\NAT$ is possible.

Coq also allows for the declaration of inductive logical properties, e.g.:
\begin{example}\small
$\Bcmd{Inductive}\:ev\!:\!\NAT\!\to\!\Bcmd{Prop}\!:=
ev_0\!:\!ev\:0\:|\:ev_{2}\!:\!\forall (n\!:\!\NAT),\:ev\:n\!\to\!ev\:(S(S\:n))$
\end{example}
It defines a family of \emph{logical types}: $ev\:0$ is a type inhabited by the term $(ev_0)$, 
$ev\:2$ is another type inhabited by $(ev_{2}\:0\:ev_0)$, and $ev\:1$ is an empty type. The 
standard interpretation is that $ev_0$ is a proof of the proposition $ev\:0$ and that there is 
no proof of $ev\:1$, that is we have $\lnot(ev\:1)$.

An intuitive interpretation of our two examples is that $\NAT$ is a set of terms, and $ev$ a 
predicate marking some of them, defining a subset of $\NAT$.

\section{Deep Embedding and Related Works}\label{sc_relworks}

\emph{Embedding} in a proof assistant consists in mechanizing a \emph{guest} logic by encoding 
its syntax and semantic into a \emph{host} logic (\cite{gor:2,bou:1,azu:1}). In a 
\emph{shallow} embedding, the encoding is partially based on a direct translation of the guest 
logic into constructs of the host logic. In a \emph{deep} embedding the syntax and the 
semantic are formalised as datatypes. At a fundamental level, taking the view presented in 
Sec. \ref{sc_introcoq}, the deep embedding of a logic is simply a definition of the set of all 
sequents (the terms) and a predicate marking those that are \emph{provable} (the inference 
rules of the guest logic being encoded as constructors of this predicate).

Shallow embeddings of \emph{B} in higher-order logics have been proposed in several papers
(cf. \cite{bod:1b,cha:1}) formalising the \emph{GSL} in \emph{PVS}, \emph{Coq} or
\emph{Isabelle/HOL}. Such embeddings are not dealing with the \emph{B} logic, and by using 
directly the host logic to express \emph{B} notions, they introduce a form of 
\emph{interpretation}. If the objective is to have an accurate formalisation of the guest 
system, the definition of a valid interpretation is difficult -- e.g. \emph{B} functions are 
relations, possibly partial or undecidable, and translating accurately this concept in 
\emph{Coq} is a tricky exercise.

\emph{BiCoq} aims at such an accurate formalisation, to pinpoint any problem of the theory 
with the objective to increase confidence in the developments when safety or security is a 
concern; in addition, we also have an implementation objective. In such cases, a deep 
embedding is fully justified -- see for example the development of a sound and complete 
theorem prover for first-order logic verified in \emph{Isabelle} proposed in \cite{rid:1}.

A deep embedding of the \emph{B} logic in \emph{Coq} is described in \cite{brk:1} (using
notations with names), to validate the \emph{base rules} used by the prover of 
\emph{Atelier-B} -- yet not checking standard \emph{B} results, and without implementation 
goal. As far as the implementation of a trusted \emph{B} prover is concerned, we can also 
mention the encoding of the \emph{B} logic as a rewriting system proposed in \cite{cir:1}.

Deep embeddings have also the advantage to clearly separate the host and the guest logics: in 
\emph{Bicoq}, excluded middle, provable in \emph{B}, is not promoted to \emph{Coq}. This 
improves readibility, and allows one to study meta-theoretical questions such as consistency. 
Furthermore, the host logic consistency is not endangered.

\section{Formalising the \emph{B} Logic in \emph{Coq}}\label{sc_formal}

In this section, we present our embedding of the \emph{B} logic in the \emph{Coq} system; the 
embedding uses a \emph{De Bruijn} representation that avoids ambiguities and constitutes an 
efficient solution w.r.t. the implementation objective (see \cite{deb:1,lia:1}). Deviations 
between \emph{B} and its formalisation are described and justified.

\begin{notation}\small\emph{B} definitions use upper case letters with standard notations.
\emph{BiCoq} uses lower case letters, and mixes \emph{B} and \emph{Coq} notations; standard 
notations are used for \emph{Coq} (e.g. $\forall$ is the universal quantification) while 
dotted notations are used for the embedded \emph{B} (e.g. $\pfor$ is the universal
quantification constructor).
\end{notation}

\begin{notation}\small$[T]$ denotes the type of the lists whose elements have type $T$.
\end{notation}

\subsection{Syntax}\label{ss_syntax}

Given a set of identifiers ($I$), the \emph{B} logic syntax defines predicates ($P$), 
expressions ($E$), sets ($S$) and variables ($V$) as follows:
\[
\small
\begin{array}{rclclclclclclclcl}
\rule{1.0cm}{0cm} &
\rule{0.7cm}{0cm} & \rule{0.8cm}{0cm} &
\rule{0.2cm}{0cm} & \rule{0.8cm}{0cm} &
\rule{0.2cm}{0cm} & \rule{0.8cm}{0cm} &
\rule{0.2cm}{0cm} & \rule{0.8cm}{0cm} &
\rule{0.2cm}{0cm} & \rule{0.8cm}{0cm} &
\rule{0.2cm}{0cm} & \rule{0.8cm}{0cm} &
\rule{0.2cm}{0cm} & \rule{0.8cm}{0cm} &
\rule{0.2cm}{0cm} & \rule{0.8cm}{0cm} \vspace{-12pt} \\
P &:=& P\!\land\!P &|& P\!\!\Rightarrow\!\!P &|& \lnot P &|& \forall\:V\cdot P
        &|& E\!=\!E &|& E\!\in\!E &|& [V\!\!:=\!\!E]P 
\\
E &:=& V &|& S &|& E\!\mapsto\!E  &|& \downarrow S &|& [V\!\!:=\!\!E]E 
\\
S &:=& \Bcmd{BIG} &|& \uparrow S &|& S\!\times\!S  &|& \{V|P\} 
\\
V &:=& I &|& V,V
\end{array}
\]
In this syntax, $[V\!:=\!E]T$ represents the (elementary) substitution, $V_1,V_2$ a list 
of variables, $E_1\!\mapsto\!E_2$ a pair of expressions, $\!\downarrow\!$ and  $\!\uparrow\!$ 
the \emph{choice} and \emph{powerset} operators, and $\Bcmd{BIG}$ a constant set. The 
comprehension set operator, while syntactically defined by $\{V|P\}$, is rejected at 
\emph{wf-checking} if not of the form $\{V|V\!\in\!S\!\land\!P\}$, with $V$ a variable not 
free in $S$·

\begin{defn}\small Other connectors are defined from the previous ones,
$P\!\!\Leftrightarrow\!\!Q$ is defined as $P\!\!\Rightarrow\!\!Q \land Q\!\!\Rightarrow\!\!P$, 
$P\!\lor\!Q$ as $\lnot P\!\!\Rightarrow\!\!Q$, and $\exists\:V\!\cdot\!P$ as 
$\lnot\forall\:V\!\cdot\!\lnot P$.
\end{defn}

The first design choice of \emph{BiCoq} is to use a pure nameless \emph{De Bruijn} notation
(see \cite{deb:1,ayd:1}), where variables are represented by indexes giving the position of 
their binder -- here the universal quantifier and the comprehension set. When an index exceeds 
the number of parent binders, it is said to be \emph{dangling} and represents a
\emph{free variable}, whose name is provided by a scope (left implicit in this paper), so that 
any syntactically correct term is semantically valid, and there is no need for well-formedness 
condition\footnote{An alternative approach to avoid well-formedness conditions is described in 
\cite{pat:1}.}. In this representation, proofs of side conditions related to name clashing are 
replaced by computations on indexes, but the index representing a variable is not constant in 
a term.

The \emph{B} syntax is formalised in \emph{Coq} by two mutually inductive types with the 
following constructors, $\Bidx$ being the set of indexes (that is, 
$\mathbb{N}\!\setminus\!\{\dbzero\}$) and $\Bnamo$ an infinite set of names with a decidable 
equality:
\[
\small
\begin{array}{rclclclclclclclcl}
\rule{1.0cm}{0cm} &
\rule{0.7cm}{0cm} & \rule{0.8cm}{0cm} &
\rule{0.2cm}{0cm} & \rule{0.8cm}{0cm} &
\rule{0.2cm}{0cm} & \rule{0.8cm}{0cm} &
\rule{0.2cm}{0cm} & \rule{0.8cm}{0cm} &
\rule{0.2cm}{0cm} & \rule{0.8cm}{0cm} &
\rule{0.2cm}{0cm} & \rule{0.8cm}{0cm} &
\rule{0.2cm}{0cm} & \rule{0.8cm}{0cm} &
\rule{0.2cm}{0cm} & \rule{0.8cm}{0cm} \vspace{-12pt} \\
\Bprd &:=& \Bprd \pand \Bprd &|& \Bprd \pimp \Bprd &|& \pnot \Bprd &|& \pfor \Bprd
        &|& \Bexp \pequ \Bexp &|& \Bexp \pins \Bexp
\\
\Bexp &:=& \evar{}\Bidx &|& \Bexp \ecpl \Bexp &|& \echs \Bexp  &|& \ebig &|& \epow \Bexp
       &|& \Bexp \epro \Bexp &|& \ecmp{\Bexp}{\Bprd} &|& \ebie{}\Bnamo
\end{array}
\]
$\Bprd$ represents \emph{B} predicates, while $\Bexp$ merges \emph{B} expressions, sets and 
variables.

Using a \emph{De Bruijn} representation, binders $\pfor$ and $\ecmp{}{}$ have no attached 
names and only bind (implicitly) a single variable. Binding over list of variables can be 
eliminated without loss of expressivity, as illustrated by the following example:

\begin{example}\label{ss_syntax_e2}\small
$\{V \:|\: V\!\in\!S_1\!\times\!S_2\!\land
              \exists V_1 \cdot (V_1\!\in\!S_1\!\land
               \exists V_2 \cdot (V_2\!\in\!S_2\!\land\!V_1\!\mapsto\!V_2\!=\!V\!\land\!P))\}$
represents
$\{V_1,V_2\:|\:V_1,V_2\!\in\!S_1\!\times\!S_2\!\land\!P\}$
\footnote{This second representation, while standard in \emph{B}, appears to be an illegal 
binding over the expression $x\!\mapsto\!y$ rather than over the variable $x,y$, but the same 
notations are used for both in \cite{abr:1} and such confusions are frequent.}
\end{example}
The constructor $\ecmp{}{}$ is further modified to be parameterised by an expression, to keep 
in the syntax definition only wf-checkable terms. Indeed, only comprehension sets of the form
$\{V\:|\:V\!\!\in\!E\land P\}$, with $V$ not free in $E$, are valid. The \emph{BiCoq} 
representation of this set is $\ecmp{e}{p}$; to reflect the non-freeness condition,
$\ecmp{e}{p}$ only binds variables in its predicate parameter $p$. By these design choices, we 
bridge the gap between syntactically correct terms and wf-checkable ones, while being 
conservative.

$\ebig$ represents the constant set $\Bcmd{BIG}$, $\evar{}$ unary (\emph{De Bruijn}) 
variables. The constructor $\ebie{}$ is without \emph{B} equivalent, and provides elements of 
$\ebig$ (cf. Par. \ref{ss_infer}).

\begin{notation}\small$\evar{i}$ denotes the application of constructor $\evar{}$ to 
$i\!:\!\Bidx$ and $\ebie{j}$ of constructor $\ebie{}$ to $j\!:\!\Bnamo$. By abuse of notation 
the variable $\evar{i}$ is also denoted simply by $i$.
\end{notation}

Finally, the elementary substitution is not considered in \emph{BiCoq} as a syntactical 
construct but is replaced by functions on terms -- substitution being introduced earlier in
\emph{B} only to be used in the description of inference rules. Note however that the full 
\emph{GSL} of \emph{B} can still be formalised by additional terms constructors (the 
\emph{explicit substitution} approach, see \cite{aba:1,cur:1}).

\begin{notation}\small$p_1\piff p_2$ is defined as $p_1\pimp p_2 \pand p_2\pimp p_1$,
$p_1\por p_2$ as $\pnot p_1 \pimp p_2$, and $\pexs p$ as $\pnot\pfor\pnot p$.
\end{notation}

\begin{notation}\small$\Btrm$ denotes the type of terms, that is the union of $\Bprd$ and 
$\Bexp$.
\end{notation}

\subsection{Dealing with the \emph{De Bruijn} Notation}\label{ss_debruijn}

\emph{De Bruijn} notation is an elegant solution to avoid complex name management, and it has
numerous merits. But it also has a big drawback, being an unusual representation for human 
readers:
\begin{example}\label{ss_debruijn_ex1}
\small If $x\!\in\!y$ is the interpretation of the term $\dbone\pins\dbtwo$, the 
interpretation of the term $\pfor\!(\dbone\pins\dbtwo)$ is $\forall t\!\cdot\!t\!\in\!x$; 
because of the binder, the scope has shifted (so $\dbtwo$ now represents $x$), and (likely) 
the semantic has been distorted.
\end{example}
In this paragraph, we illustrate some of the consequences of using a \emph{De Bruijn} notation,
as well as how to mask such consequences from the users.

\subsubsection{Induction} When defining type $\Btrm$, \emph{Coq} automatically generates the 
associated structural induction principle. As illustrated in Ex. \ref{ss_debruijn_ex1}, it is 
however not semantically adequate, because it does not reflect \emph{De Bruijn} indexes 
scoping. A more interesting principle is derived in \emph{BiCoq} by using the syntactical 
depth function $\depth$ of a term as a well-founded measure:
\[
\small
\begin{array}{l}
\vspace{-16pt}\\
\forall\:(P\!:\!\Btrm\!\to\!Prop),
(\forall\:(t\!:\!\Btrm),(\forall\:(t'\!:\!\Btrm),
                               \depth(t')\!<\!\depth(t) \to P\:t') \to P\:t) \to
\forall\:(t\!:\!\Btrm),\:P\:t
\end{array}
\]
With this principle, for the term $\pfor\!(\dbone\pins\dbthree)$ (that is,
$\forall t\!\cdot\!t\!\in\!y$) we can choose to use an induction hypothesis on 
$\dbone\pins\dbtwo$ (that is, $x\!\in\!y$) instead of $\dbone\pins\dbthree$ (that is, 
$x\!\in\!z$).

\subsubsection{Non-Freeness} The \emph{B} notation $V \backslash T$ means that the variable 
$V$ does not appear free in $T$. Non-freeness is defined in \emph{BiCoq} as a type 
$\Bnotfree\!:\!\Bidx\!\to\!\Btrm\!\to\!Prop$ (a relation between $\Bidx$, representing the 
variables, and $\Btrm$), with the following rules\footnote{The rules for the other 
constructors are trivial and can be obtained by straightforward extension, e.g. here
$i \Bnotfree p$ and $i \Bnotfree q$ allow to derive $i \Bnotfree p\pimp q$.}:
\[
\small
\begin{array}{ccccc}
\vspace{-16pt} \\
\begin{array}{c} \\\hline \rule{0cm}{12pt} i \Bnotfree \ebig \end{array}
\quad&\quad
\begin{array}{c} \\\hline \rule{0cm}{12pt}  i \Bnotfree \ebie{k}\end{array}
\quad&\quad
\begin{array}{c} i_1 \neq i_2 \\\hline \rule{0cm}{12pt}  i_1 \Bnotfree i_2\end{array} 
\quad&\quad
\begin{array}{c} (i\!+\!1)\Bnotfree p\\\hline \rule{0cm}{12pt} i\Bnotfree\pfor p \end{array}
\quad&\quad
\begin{array}{c} i \Bnotfree e \quad (i\!+\!1) \Bnotfree p\\\hline \rule{0cm}{12pt}
                 i \Bnotfree \ecmp{e}{p} \end{array}\\
\end{array}
\]
The two first rules are axioms, the associated constructors are atomic and do not interact 
with variables. The rules for $\pfor$ and $\ecmp{}{}$ reflect the fact that the associated 
constructors are binders and therefore shift the scope.

\subsubsection{Binding, Instantiation and Substitution} It is possible to define functions to 
simulate \emph{B} binding (that is the use of $\forall$ or $\{\}$, representing 
$\lambda$-abstraction). These functions constitute a built-in user interface to produce 
\emph{De Bruijn} terms while using the usual representation, making \emph{De Bruijn} indexes 
and their management invisible to the user (see also \cite{gor:1} for a similar approach):
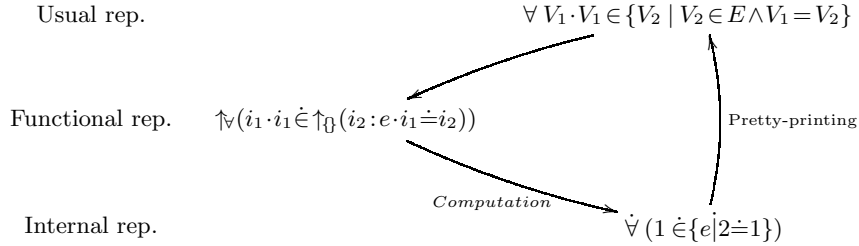
\[
\small
\xymatrix{
\text{Usual rep.} & & \!\!\!\!\!\!\!\!\forall\:V_1\!\cdot\!V_1\!\in\!
  \{V_2\:|\:V_2\!\in\!E\!\land\!V_1\!=\!V_2\}
 \ar@/_/@{->}[dl]  \ar@/^/@{<-}[dd]^{\text{Pretty-printing}}\\
\text{Functional rep.} &
\!\!\!\!\!\!\!\!\Bbdfor{i_1}{i_1\pins\Bbdcmp{i_2}{e}{i_1\pequ i_2}}
\ar@/_/@{->}[dr]_{Computation\quad}\\
\text{Internal rep.} & & \pfor(\dbone\:\pins\ecmp{e}{\dbtwo\pequ\dbone})\\
}
\]
The binding functions are defined by:
\[
\small
\begin{array}{lll}
\vspace{-16pt}\\
\Bbdfor{i}{p}\!:=\!\pfor \Bcmd{Bind}\:i\:\dbone\:p \quad & \quad
\Bbdcmp{i}{e}{p}\!:=\!\ecmp{e}{\Bcmd{Bind}\:i\:\dbone\:p} \quad & \quad
\Bbdexs{i}{p}\!:=\!\pexs \Bcmd{Bind}\:i\:\dbone\:p \\
\multicolumn{3}{l}{
\begin{array}{lcl}
\vspace{-6pt}\\
\multicolumn{3}{l}{\Bcmd{Bind}(i_1\:i_2\!:\!\Bidx)(t\!:\!\Btrm)\!:\!\Btrm\!:=\!match\:t\:with}
\\
\quad | \quad \ebig \quad|\quad \ebie{j'} & \Rightarrow & t\\
\quad | \quad \evar{i'} & \Rightarrow &
 t\text{ if }i'\!<\!i_2\text{, or else }
 \evar{i_2}\text{ if }i'\!=\!i_1\text{, or else }\evar{i'\!+\!1}\\
\quad | \quad \pfor p' & \Rightarrow & \pfor (\Bcmd{Bind}\:(i_1\!+\!1)\:(i_2\!+\!1)\:p')\\
\quad | \quad \ecmp{e'}{p'} & \Rightarrow & \ecmp {\Bcmd{Bind}\:i_1\:i_2\:e'\:}
                                                {\:\Bcmd{Bind}\:(i_1\!+\!1)\:(i_2\!+\!1)\:p'}\\
\quad | \quad \ldots & \Rightarrow & \ldots \text{ (straightforward extension)}\\
\end{array}
}
\end{array}
\]
On the same principles, the definition of instantiation functions (for elimination of 
$\forall$ or $\{\}$, representing $\beta$-reduction and denoted by 
$\Binfor{e}{p}\!:\!\Bprd\!\to\!\Bexp\!\to\!\Bprd$ and
$\Bincmp{e_2}{e_1}\!:\!\Bexp\!\to\!\Bexp\!\to\!\Bprd$) is straightforward -- being partial,
these functions just require in \emph{Coq} an additional proof parameter (omitted in this 
paper) that the term is of the expected form. Finally, it is also possible to define a 
substitution function\footnote{Substitution and instantiation may seem similar in usual 
notation, but their differences are emphasised when using \emph{De Bruijn} notation.}:
\[
\small
\begin{array}{lcl}
\vspace{-16pt}\\
\multicolumn{3}{l}
{\Baffec{i}{e}{t}\!:\!\Bidx\!\to\!\Bexp\!\to\!\Btrm\!\to\!\Btrm\!:=\!match\:t\:with}\\
\quad | \quad \ebig\quad|\quad\ebie{j'} & \Rightarrow & t\\
\quad | \quad \evar{i'} & \Rightarrow & if\:i'\!=\!i\:then\:e\:else\:t\\
\quad | \quad \pfor p' & \Rightarrow & \pfor \Baffec{i\!+\!1}{\Bcmd{Lift}(e)}{p'}\\
\quad | \quad \ecmp{e'}{p'} & \Rightarrow & \ecmp{\Baffec{i}{e}{e'}}
                                               {\Baffec{i\!+\!1}{\Bcmd{Lift}(e)}{p'}}\\
\quad | \quad \ldots & \Rightarrow & \ldots \text{ (straightforward extension)}\\
\end{array}
\]
where $\Bcmd{Lift}$, not detailed in this paper, increments dangling \emph{De Bruijn} indexes.
Remember that substitution is introduced early in \emph{B} as a syntactical construct, but 
only to be used in inference rules. We consider that such rules are better represented using 
the resulting term (that is, the reduction of the application of the substitution).

Once these functions are defined, numerous lemmas are proved, such as the (in)famous ones 
describing all possible interactions between lifting, binding, instantiation and substitution.
The following results are then derived, proving the irrelevance of $\alpha$-renaming or 
describing relationships between instantiation, binding and substitution (with $=$ the 
\emph{Coq} term structural equality):
\[
\small
\begin{array}{ll}
\rule{5.7cm}{0pt} & \rule{5.7cm}{0pt} \vspace{-16pt} \\
i_2 \Bnotfree p\!\to\:\:
\Bbdfor{i_1}{p}\!=\:\Bbdfor{i_2}{\Baffec{i_1}{i_2}{p}}
&
i_2 \Bnotfree p\!\to\:
 \:\Bbdcmp{i_1}{e}{p}\!=\:\Bbdcmp{i_2}{e}{\Baffec{i_1}{i_2}{p}}\\
\Binfor{i}{\Bbdfor{i}{p}}\!=\!p
&
\Bincmp{i}{\Bbdcmp{i}{e}{p}}\!=\!i\pins e\:\pand\:p\\
\Binfor{e}{\Bbdfor{i}{p}}\!=\!\Baffec{i}{e}{p}\\
\end{array}
\]

\subsection{Inference Rules}\label{ss_infer}

\begin{table}
\caption{Encoding of the \emph{B} inference rules}
\label{BINF}
\[
\small
\begin{array}{|l|l|}
\hline
\rule{0cm}{12pt}\text{\emph{B} inference rules}
&
\text{\emph{BiCoq} formalisation}
\\
\hline

\begin{array}{c}
\\\hline P \vdash P
\end{array}
&
\text{None, derived from }\rulename{\in}

\\

\begin{array}{c}
P \text{ appears in } \Gamma \\\hline \Gamma \vdash P
\end{array}
&
p\!\in\!g \to g \Binf p\rulename{\in}

\\

\begin{array}{c}
\Gamma' \text{ includes } \Gamma \quad\quad \Gamma \vdash P
\\\hline \Gamma' \vdash P
\end{array}
&
g_1\Binf p \to g_1\!\subseteq\!g_2 \to g_2\Binf p\rulename{\subseteq}

\\

\begin{array}{c}
\Gamma \vdash P \quad \Gamma,P \vdash Q \\\hline \Gamma \vdash Q
\end{array}
&
\text{None, derived from }\rulename{\lnot_{n}}\rulename{\lnot_{p}}
                          \rulename{\subseteq}\rulename{\in}

\\

\begin{array}{c}
\Gamma \vdash P \Rightarrow Q \\\hline \Gamma,P\vdash Q
\end{array}
\quad
\begin{array}{c}
\Gamma,P\vdash Q \\\hline \Gamma \vdash P \Rightarrow Q
\end{array}
&
\begin{array}{l}
\!g\Binf p_1 \pimp p_2 \to g,p_1\Binf p_2 \\
\!g,p_1\Binf p_2 \to g\Binf p_1 \pimp p_2
\end{array}

\\

\begin{array}{c}
\Gamma \vdash P \quad \Gamma \vdash Q \\\hline \Gamma \vdash P \land Q
\end{array}
&
g\Binf p_1 \to g\Binf p_2 \to g\Binf p_1 \pand p_2 \rulename{\land_i}

\\

\begin{array}{c}
\Gamma \vdash P \land Q \\\hline \Gamma \vdash P
\end{array}
\quad
\begin{array}{c}
\Gamma \vdash P \land Q \\\hline \Gamma \vdash Q
\end{array}
&
\begin{array}{l}
\!g\Binf p_1 \pand p_2 \to g\Binf p_1\\
\!g\Binf p_1 \pand p_2 \to g\Binf p_2
\end{array}

\\

\begin{array}{c}
\Gamma,Q \vdash P \quad \Gamma,Q \vdash \lnot P \\\hline \Gamma \vdash \lnot Q
\end{array}
&
g,p_2\Binf p_1\to g,p_2\Binf\pnot p_1\to g\Binf\pnot p_2 \rulename{\lnot_p}

\\

\begin{array}{c}
\Gamma,\lnot Q \vdash P \quad \Gamma,\lnot Q \vdash \lnot P \\\hline \Gamma \vdash Q
\end{array}
&
g,\pnot p_2\Binf p_1\to g,\pnot p_2\Binf\pnot p_1\to g\Binf p_2 \rulename{\lnot_n}

\\

\begin{array}{c}
\\\hline \Gamma \vdash E=E
\end{array}
&
g \Binf e \pequ e

\\

\begin{array}{c}
\Gamma \vdash P \quad V \backslash \Gamma \\\hline
\Gamma \vdash \forall\: V \cdot P
\end{array}
&
i \Bnotfree g \to g \Binf p \to g \Binf \Bbdfor{i}{p} \rulename{\forall_i}

\\

\begin{array}{c}
\Gamma \vdash \forall\: V \cdot P\\\hline \Gamma \vdash [V:=E]P
\end{array}
&
g \Binf \Bbdfor{i}{p} \to g \Binf \Baffec{i}{e}{p}

\\

\begin{array}{c}
V \backslash S
\\\hline \vdash E\!\in\!\{V|V\!\!\in\!\!S\!\land\!P\}\!\Leftrightarrow\!E\!\!\in\!\!S\!\land\!
 [V\!\!:=\!\!E]P
\end{array}
&
\Binf e_1\pins\Bbdcmp{i}{e_2}{p}\piff e_1\pins e_2\pand \Baffec{i}{e_1}{p}

\\

\begin{array}{c}
\Gamma \vdash E=F \quad \Gamma \vdash [V\!\!:=\!\!E]P\\\hline \Gamma \vdash [V\!\!:=\!\!F]P
\end{array}
&
g\Binf e_1\pequ e_2\to g\Binf\Baffec{i}{e_1}{p}\to g\Binf\Baffec{i}{e_2}{p}

\\

\begin{array}{c}
V \backslash S
\\\hline
\vdash \exists\:  V \cdot (V\!\in\!S) \Rightarrow \downarrow\! S\!\in\!S
\end{array}
&
i \Bnotfree e \to g\Binf\Bbdexs{i}{i\pins e}\pimp\echs e\pins e

\\

\begin{array}{c}
V \backslash S,\!T \\\hline
\vdash S\!\in \uparrow\! T \Leftrightarrow \forall\: V \cdot (V\!\in\!S \Rightarrow V\!\in\!T)
\end{array}
&
i \Bnotfree e_1 \to i \Bnotfree e_2 \to
g \Binf e_1 \pins \epow e_2 \piff \Bbdfor{i}{i\pins e_1\pimp i\pins e_2}

\\

\begin{array}{c}
V \backslash S,\!T \\\hline
\vdash
\left(
\begin{array}{l}
\:\:\: \forall\:V\!\cdot\!(V\!\in\!S\!\Rightarrow\!V\!\in\!T) \\
\land \forall\:V\!\cdot\!(V\!\in\!T\!\Rightarrow\!V\!\in\!S)
\end{array}
\right)
\!\Leftrightarrow\!S\!=\!T
\end{array}
&
\begin{array}{l}
\!g\Binf e_1\pins\epow e_2\to g\Binf e_2\pins\epow e_1\to g\Binf e_1\pequ e_2
\end{array}

\\

\begin{array}{c}
\\\hline
\vdash \Bcmd{infinite}(BIG)
\end{array}
&
\begin{array}{ll}
\!g\Binf\ebie{j}\pins\ebig\\
\!j_1\neq j_2\to g\Binf\pnot(\ebie{j_1}\pequ\ebie{j_2})\\
\end{array}

\\

\begin{array}{c}
\\\hline
\vdash (E\!\!\mapsto\!\!F)\!\in\!(S\!\!\times\!\!T) \Leftrightarrow 
(E\!\!\in\!\!S)\!\land\!(F\!\!\in\!\!T)
\end{array}
&
\begin{array}{l}
\rule{0cm}{15pt}
\!g\Binf e_1\ecpl e_2\pequ e_3\ecpl e_4 \to g\Binf e_1\pequ e_3\\
\!g\Binf e_1\ecpl e_2\pequ e_3\ecpl e_4 \to g\Binf e_2\pequ e_4\\
\!\begin{array}{l}
\!i_1 \Bnotfree e\pins(e_1\epro e_2) \to i_2 \Bnotfree e\pins(e_1\epro e_2) \to
 i_1 \neq i_2 \to\\
  \: g\Binf\!\!\Bbdexs{i_1}{i_1\pins e_1\pand
    \!\Bbdexs{i_2}{i_2\pins e_2\pand e\pequ i_1\ecpl i_2}}\piff e\pins(e_1\epro e_2)
\end{array}
\end{array}

\\\hline
\end{array}
\]
\end{table}

Having formalised the \emph{B} syntax and defined some functions and properties on terms, the 
next step is to encode the \emph{B} inference rules. Thanks to the use of the functional 
representation described in the previous paragraph, \emph{BiCoq} rules look very much like the 
standard \emph{B} rules. The translation is therefore straightforward, merely a syntactical 
one, and the risk of error is very limited.

In our formalisation sets of hypothesis are represented by lists, with membership ($\in$) and 
inclusion ($\subseteq$) as well as the pointwise extension of non-freeness ($\Bnotfree$).
The \emph{B} inference rules are formalised as constructors of an inductive type
$\Binf\!:\![\Bprd]\!\to\!\Bprd\!\to\!Prop$, that is $g \Binf p$ is the \emph{Coq} type of all 
\emph{B} proofs of $p$ under the assumptions $g$. Such a type may be inhabited (i.e. $p$ is 
provable assuming $g$) or empty (i.e. there is no proof of $p$ under the assumptions of $g$).

The \emph{B} rules and their encoding as constructors are detailed in Tab.
\ref{BINF}, universal quantifications being omitted (the types are $g,g_1,g_2\!:\![\Bprd]$; 
$p,p_1,p_2\!:\!\Bprd$; $e,e_1,e_2,e_3,e_4\!:\!\Bexp$, $i,i_1,i_2\!:\!\Bidx$ and 
$j,j_1,j_2\!:\!\Bnamo$). For most of them, translation is straightforward, only taking care to 
use functional substitution and binding where appropriate. On the other hand, the use of the 
functional representation imposes to keep the syntactical side conditions, except for the 
comprehension set rule, where such condition is embedded in the syntax; new rules have to be 
derived to benefit of the internal \emph{De Bruijn} representation.

Only the last two \emph{B} inference rules deserve discussion. The first one of these
indicates that the constant set $\Bcmd{BIG}$ is infinite, using the $\Bcmd{infinite}$ 
\emph{B} predicate defined by a fixpoint; unfolding this definition to produce a translation is
possible, but not practical. Therefore, this rule is replaced in \emph{BiCoq} by two different 
rules allowing to exhibit an infinity of elements of $\Bcmd{BIG}$, $\Bnamo$ being itself 
infinite.

The last rule, defining the semantics of pairs and products, is more interesting. A 
straightforward translation of this rule indeed leads to the impossibility to prove, in 
\emph{BiCoq}, the following theorems from \cite{abr:1}:
\[
\small
\begin{array}{l}
\vspace{-16pt}\\
\vdash (E\!\mapsto\!F)\!=\!(E'\!\mapsto\!F') \Rightarrow E\!=\!E' \land F\!=\!F'\\
\vdash S\in\uparrow\!U \land T\in\uparrow\!V\Rightarrow(S \times T)\in\uparrow\!(U \times V)\\
\end{array}
\]
The proof of the first result provided in \cite{abr:1} is flawed, due to a confusion between 
pairs of expressions and lists of variables (as pointed out in \cite{mus:1}), both using the 
same notation -- and cannot be corrected in the absence of a form of destructor for pairs. On 
the other hand, the proof of the monotonicity of cartesian product w.r.t. inclusion is not 
detailed in \cite{abr:1}, being considered trivial. However, using the listed rules, one may 
derive predicates of the form $V\!\in\!S\!\times\!T$ but without being able to constraint $V$ 
to be a pair to apply the last rule (a classical problem of the untyped $\lambda$-calculus). 
Basically, injectivity and surjectivity rules are lacking; these observations, probably well 
known of the \emph{B} gurus but not documented to our knowledge, have led us to replace this 
\emph{B} rule by three new rules in order to be able to prove the expected theorems. Again, 
this process illustrates our conservative approach.

\section{Proofs in \emph{BiCoq}}\label{sc_proofs}

\subsection{Standard \emph{B} Proofs}\label{ss_classic}

Using the definition of $\Binf$, we formally prove in \emph{BiCoq} all propositional calculus 
and predicate calculus results of \cite{abr:1}, using the functional representation and 
following the proposed proof structure, e.g.:
\[
\small
\begin{array}{l}
\vspace{-16pt}\\
i_1\Bnotfree g\!\to\!i_1\Bnotfree p\!\to\!g\Binf\Baffec{i_2}{i_1}{p}\!\to\! 
 g\Binf\!\Bbdfor{i_2}{p}
\text{, that is }
\begin{array}{c}
 \Gamma\vdash[V_2\!:=\!V_1]P\quad V_1\backslash\Gamma,\!P\\\hline
 \Gamma\vdash\forall\:V_2\!\cdot\!P
\end{array}
\end{array}
\]

To assist the proof construction \emph{BiCoq} provides \emph{Coq} tactics written in the 
\emph{Coq} tactic language \cite{del:1}. For example, the propositional calculus procedure 
described in \cite{abr:1}, proposing a strategy based on propositional calculus theorems, is 
provided as a \emph{Coq} tactic. More technical \emph{Coq} tactics are also available in 
\emph{BiCoq}, e.g. to obtain proved fresh variables.

An alternative form of theorems is also derived, using the internal \emph{De Bruijn} 
representation; e.g. the $\pfor\!$-introduction rule (to be compared with 
$\rulename{\forall_i}$) is:
\[
\small
\begin{array}{l}
\vspace{-16pt}\\
i\Bnotfree g\to i\Bnotfree\pfor p\to g\Binf\Bcmd{Inst}\:i\:\dbone\:p\to g\Binf\pfor p
\end{array}
\]
These last results are of course rather technical, not benefiting from the functional 
representation. Yet they have some interest, for technical lemmas or as derived rules in which 
only semantical side conditions remain (computations over \emph{De Bruijn} indexes dealing 
with the syntactical ones).

\subsection{Mixing \emph{BiCoq} and \emph{Coq} Logics}\label{ss_mixing}

As it is standard in such a deep embedding (e.g. see \cite{brk:1}), \emph{BiCoq} provides also 
results expressing relations between host and guest logics:
\[
\small
\begin{array}{ll}
\rule{5.7cm}{0pt} & \rule{5.7cm}{0pt} \vspace{-16pt} \\
(g\Binf p\lor g\Binf q)\!\to\!g\Binf p\dot{\lor}q
&
g\Binf p\pimp g\!\to\!(g\Binf p\to g\Binf p)
\\
(g\Binf p\land g\Binf q)\!\leftrightarrow\!g\Binf p\pand q
&
(\forall\:(y\!:\!\Bidx),\:g\Binf\Baffec{x}{y}{p})\!\leftrightarrow\!g\Binf\Bbdfor{x}{p}
\end{array}
\]
Asymmetrical results mark the differences between the classical \emph{B} logic and the 
constructive \emph{Coq} logic -- e.g. a reciprocal of the first rule, combined with the 
excluded middle, would prove that for any predicate $p$ either $\Binf p$ or $\Binf \pnot p$, 
which of course is not the case. This emphasises the fact that both logics are well separated, 
the \emph{B} logic being embedded has an external theory.

By providing the best of both worlds, these results constitute efficient proof tactics. For 
example, the last theorem does not reflect non-freeness side conditions from \emph{B} to the 
\emph{Coq} logic (\emph{Coq} taking care of such conditions automatically).

\section{Developing a Proved \emph{B} Toolkit}\label{sc_environ}

In this section, we detail how \emph{BiCoq} is used as a framework for the development of 
formally checked \emph{B} toolkits. \emph{Coq} offers mechanisms to extract programs from 
constructive proofs (i.e. software from logical definitions and theorems), but a different 
approach is chosen here. Indeed, \emph{BiCoq} includes code (in the form of functions using 
the \emph{ML}-like internal language of \emph{Coq}) which is proved correct. This code is 
extractible by a pure syntactical process, e.g. in \emph{Objective Caml}, using the extraction 
mechanism of \emph{Coq}. By doing so, we obtain proved \emph{B} tools whose code is small, 
readable and efficient -- and independent of \emph{Coq}.

\begin{notation}\small$\Bbol$ represents the booleans, $\top$ being $true$ and $\bot$ being 
$false$.
\end{notation}

\begin{notation}\small Hat notations are used for boolean functions (e.g. $\widehat{\land}$ 
is the boolean and).
\end{notation}

\subsection{Implementing Decidable Properties}\label{ss_implem}

For $P$ and $f$ respectively a predicate and a boolean function over a type $S$, we note
$(P\!\rightsquigarrow\!f)$ when $f$ decides $P$, i.e. when the following property is proved:
\[
\small
\begin{array}{l}
\vspace{-16pt}\\
\forall(s\!:\!S), (f(s)\!=\!\top \to P(s)) \land (f(s)\!=\!\bot \to \lnot P(s))
\end{array}
\]
By defining folding as the extension of predicates and functions to lists, we prove that if 
$f$ decides $P$, then the folding of $f$  decides the folding of $P$:
\[
\small
\begin{array}{l}
\vspace{-16pt}\\
\Bcmd{Fold_p}(P)\!:=\!fun (L\!:\![S]) \Rightarrow \forall (s\!:\!S), s\!\in\!L\!\to\!P(s)\\
\Bcmd{Fold_f}(f)\!:=\!fun (L\!:\![S]) \Rightarrow
if\:\Bcmd{empty}(L)\:then\:\top\:else\:
f(\Bcmd{head}(L)) \widehat{\land} \Bcmd{Fold_f}(f)(\Bcmd{tail}(L))\\
(P\!\rightsquigarrow\!f) \to (\Bcmd{Fold_p}(P)\!\rightsquigarrow\!\Bcmd{Fold_f}(f))
\end{array}
\]

\begin{example}[Non-freeness]\small Non-freeness is defined in \emph{B} as a logical 
proposition and represented by the inductive type $\Bnotfree$ in \emph{BiCoq}. Our 
implementation strategy consists in developing a program 
$\widehat{\smallsetminus}\!:\!\Bidx\!\to\!\Btrm\!\to\!\Bbol$ and to prove that 
$(\Bnotfree\!\rightsquigarrow\!\widehat{\smallsetminus})$. Hence $\widehat{\smallsetminus}$ 
and its extension (checking that a variable does not occur free in a list of hypotheses) are 
proved correct and can be extracted.
\end{example}

In \emph{BiCoq} this approach is systematic; all typed equalities are implemented and proved 
correct (e.g. term equality), as well as non-freeness, list membership, inclusion, etc. to 
constitute our formally checked \emph{B} toolkit.

\subsection{A Proved Prover for the \emph{B} Logic}\label{ss_prov}

In this paragraph we focus on the definition of an extractible prover to conduct first-order 
\emph{B} proofs for standard \emph{B} developments.

\emph{BiCoq} includes programs, named \emph{B tactics} in the following, to simulate the 
application of \emph{B} inference rules or theorems. By providing such a dedicated piece of 
code for each of the inference rules listed in Tab. \ref{BINF}, and by proving them correct, 
we got a correct and complete prover (that is, any standard \emph{B} result can be derived 
using this prover). 

To this end, a type for \emph{sequents} is defined as the product $[\Bprd]\!\times\!\Bprd$; 
for $g\!:\![\Bprd]$ and $p\!:\!\Bprd$ we denote $g\!\Vdash\!p$ the associated pair. While
$g \Binf p$ is the type of \emph{B} proofs of $p$ under the assumptions $g$, that can be 
inhabited or not, $g \Vdash p$ is a syntactical construct extending $\Btrm$. To interpret a 
sequent, we use the translation $\Bcmd{Trans_\vdash}$ that for a pair $g\!\Vdash\!p$ returns 
the type $g \Binf p$ (and its extension derived by $\Bcmd{Fold_p}$).

A \emph{B} tactic is a function $T§_B\!:\Vdash\to\![\Vdash]$ that, provided a goal 
$g\!\Vdash\!p$, returns a list of subgoals $[g_1\!\Vdash\!p_1,\ldots,g_n\!\Vdash\!p_n]$ which 
together are sufficient to prove $g\!\Vdash\!p$; if a \emph{B} tactic concludes (proves the 
goal) this list is empty. The following (elementary) examples give the definition of the 
\emph{B} tactics associated respectively to the inference rules $\rulename{\in}$ and 
$\rulename{\land_i}$:
 \begin{example}\small
$\Bcmd{T_{\in}}(s)\!:=\!let\:(g\!\Vdash\!p\!:=\!s)\:in\:
  i\:\!\!f\:p\widehat{\in}g\:then\:[]\:else\:[s]$
\end{example}
\begin{example}\small
$\Bcmd{T_{\land_i}}(s)\!:=\!let\:(g\!\Vdash\!p\!:=\!s)\:in\:
  match\:p\:with\:p_1 \pand p_2 \Rightarrow [g\!\Vdash\!p_1,g\!\Vdash\!p_2]\:|
                \:\_ \Rightarrow [s]$
\end{example}
The implementation strategy described in Par. \ref{ss_implem} is now particularly relevant, as 
$\Bcmd{T_{\in}}$ uses the boolean function $\widehat{\in}$ instead of the logical proposition 
$\in$.

Following the same principles, numerous (much more complex) \emph{B} tactics are provided in 
\emph{BiCoq}, implementing theorems or strategies, such as the decision procedure for 
propositional calculus described in \cite{abr:1}. For each \emph{B} tactic $T_B$, the 
correctness is ensured by a proof of the following property:
\[
\small
\begin{array}{l}
\vspace{-16pt}\\
\forall\:(s:\Vdash), \Bcmd{Trans_\vdash}(T_B(s))\to\Bcmd{Trans_\vdash}(s)
\text{, that is }
\begin{array}{c}
g_1 \vdash p_1\quad\ldots\quad g_n \vdash p_n\\\hline g \vdash p
\end{array}
\end{array}
\]

Thanks to the functions defined in Par. \ref{ss_debruijn}, management of the \emph{De Bruijn}
indexes can be hidden from the users of the \emph{B} tactics. With the programs already 
provided in \emph{BiCoq} (such as non-freeness, binding, etc.), these \emph{B} tactics 
constitute the core of a proved prover. This prover still lacks automation and \emph{HMI}, and 
should be coupled with other tools, for example a \emph{B} parser using the platform \emph{BRILLANT} \cite{sco:1}.

\section{Higher-Order Considerations and Extensions}\label{sc_extend}

While the \emph{B} logic is first-order, various definitions and proofs in \cite{abr:1} are 
conducted in a higher-order meta-logic: results in propositional calculus are proved by 
induction over terms, and refinement is defined by quantification over predicates before being 
transformed into an equivalent first-order definition. Using the higher-order framework
provided by \emph{Coq}, \emph{BiCoq} can clearly be extended to integrate and to formally check
such concepts.

New results can also be derived; for example, using the proof depth function
$\depth_{\vdash}\!:\!\Binf\!\!\to\!\Bidx$, we obtain a depth induction principle on \emph{B} 
proof trees e.g. for results about proof rewriting. Other results, proved in higher-order 
logic, are applicable in first-order \emph{B} logic, and implemented as \emph{B} tactics for 
standard \emph{B} proofs. This is the case for the following congruence results.

\subsubsection{Predicate Substitution.} We extend the \emph{B} logic syntax with a new 
\emph{predicate variable} constructor $\ppvr{}\Bnamp\!:\!\Bprd$ ($\Bnamp$ being an infinite 
set of names with a decidable equality), without adding any inference rules in order not to 
enrich the \emph{BiCoq} logic\footnote{However, some new (propositional) sequents became 
provable, such as ${\ppvr{k}}\Binf{\ppvr{k}}$.}. Only limited modifications of \emph{BiCoq} 
are required to deal with this new constructor, e.g. non-freeness with the additional rule
$\forall\:(i\!:\!\Bidx)(k\!:\!\Bnamp),\:i\Bnotfree\ppvr{k}$.

Predicate variables play a role similar to the one of the variables -- they are placeholders 
that can be replaced by a predicate using the substitution function
$\Baffprd{k}{p_1}{p_2}\!:\!\Bnamp\!\to\!\Bprd\!\to\!\Bprd\!\to\!\Bprd$, not detailed in this 
paper, that mimicks the expression substitution function (see Par. \ref{ss_debruijn}). Thanks 
to this extension, we can prove the following congruence rules for $\piff$ and implement 
associated \emph{B} tactics that can be used e.g. to unfold a definition in a term, even under 
binders:
\[
\small
\begin{array}{ll}
\vspace{-16pt}\\
g\Binf p_1\piff p_2\!\to\!g\Binf \Baffprd{j}{p_1}{p} \piff \Baffprd{j}{p_2}{p}
\:\: & \:\:
g\Binf p_1\piff p_2\!\to\!g\Binf \Baffprd{j}{p_1}{e} \pequ \Baffprd{j}{p_2}{e}
\end{array}
\]

\begin{example}\label{sc_extend_ex1}\small
$x\pequ 0,y\pins\NAT\Binf y\!\leq\!x \piff y \pequ 0$, therefore we immediately derive (in one 
step)
$x\pequ 0,y\pins\NAT\Binf \Bbdfor{v}{v \pins \Bbdcmp{t}{\NAT}{t\!\leq\!y\pand y\!\leq\!x}}\piff
\Bbdfor{v}{v \pins \Bbdcmp{t}{\NAT}{t\!\leq\!y \pand y\pequ 0}}$
\end{example}
Note that predicate substitution and expression substitution mechanically forbid the capture 
of variables in the substituted subterm, by lifting dangling \emph{De Bruijn} indexes when 
crossing a binder. That is, in Ex. \ref{sc_extend_ex1}, if $v$ or $t$ appear free in the 
substituted subterm, they escape capture during substitution.

\subsubsection{Predicate Grafting.} Other congruence results can be derived for 
\emph{grafting} of predicates, a modified substitution (not lifting the substituted subterm) 
allowing for the capture of variables:
\[
\small
\begin{array}{lcl}
\vspace{-16pt}\\
\multicolumn{3}{l}
{\Bgraftprd{k}{p}{t}\!:\!\Bnamp\!\to\!\Bprd\!\to\!\Btrm\!\to\!\Btrm\!:=\!match\:t\:with}\\
\quad | \quad \ebig\quad|\quad\ebie{j'}\quad|\quad\evar{i'} & \Rightarrow & t\\
\quad | \quad \ppvr{k'} & \Rightarrow & if\:k'\!=\!k\:then\:p\:else\:t\\
\quad | \quad \pfor p' & \Rightarrow & \pfor \Bgraftprd{k}{p}{p'}\\
\quad | \quad \ecmp{e'}{p'} & \Rightarrow & \ecmp{\Bgraftprd{k}{p}{e'}}
                                               {\Bgraftprd{k}{p}{p'}}\\
\quad | \quad \ldots & \Rightarrow & \ldots \text{ (straightforward extension)}\\
\end{array}
\]
The associated congruence results and proofs are technical, and not detailed in this paper. We 
just provide for illustration a simplified version of these results:
\[
\small
\begin{array}{ll}
\vspace{-16pt}\\
\Binf p_1\piff p_2 \to g\Binf \Bgraftprd{j}{p_1}{p} \piff \Bgraftprd{j}{p_2}{p}
\quad & \quad
\Binf p_1\piff p_2 \to g\Binf \Bgraftprd{j}{p_1}{e} \pequ \Bgraftprd{j}{p_2}{e}
\end{array}
\]

\begin{example}\label{sc_extend_ex2}\small
$g\Binf\Bgraftprd{k}{\pnot\pnot p}{q}\piff\Bgraftprd{j}{p}{q}$, 
that is the elimination of double negations in a subterm (even if dangling \emph{De Bruijn} 
indexes of $p$ are bound in $q$)
\end{example}

\subsubsection{Remark.} Results such as the ones in Exs. \ref{sc_extend_ex1} or 
\ref{sc_extend_ex2} are provable in \emph{B}, on a case-by-case basis, with a first-order 
proof depending on the structure of the term in which substitution or grafting is done. It is 
therefore conceivable to develop a specific (and likely complex) \emph{B} tactic automatically 
building for such goals a proof using the \emph{B} inference rules. On the contrary, the 
proposed extensions provide a new approach through results derived from a higher-order proof; 
the associated \emph{B} tactics are therefore simpler, and produce generic (and shorter) 
proofs by using not only the \emph{B} inference rules but also induction on $\Btrm$.

\section{Conclusion}\label{sc_conc}

Through an accurate deep embedding of the \emph{B} logic in \emph{Coq}, we identify shortfalls 
or confusions in \cite{abr:1} and propose amendments in order to be able to validate standard
results -- improving the confidence in the method and in the developments conducted with it.
We describe a strategy to further benefit from this deep embedding by implementing verified 
\emph{B} tools, extractible to be used independently of \emph{Coq}. The approach is 
illustrated by the development of \emph{B} tactics that constitute a complete and correct 
prover -- usable to conduct proofs (provided further automation), or to check proofs produced 
by other tools. The objective, again, is to have better confidence in the developments 
conducted in \emph{B}.

We also explain how, benefiting from the higher-order features of \emph{Coq}, new results for 
\emph{B} can be derived, and present an extension to derive congruence theorems related to 
equivalence, implemented in our prover.

All the results presented in this paper are mechanically checked; \emph{BiCoq} currently 
represents about 550 definitions (i.e. types, properties, functions), 750 theorems and proofs 
in \emph{Coq} -- and about 6 man.months of development. It has now to be extended with the 
following definitions and results:
\begin{itemize}
\item Generation by the prover of \emph{B} proof terms checkable by \emph{Coq}.
\item Use of a locally nameless \emph{De Bruijn} representation with named free variables to 
derive unified congruence results (merging substitution and grafting).
\item Fixpoint constructs, with application to the definition of natural numbers in the 
\emph{B} style; on the innovative side, we expect to derive inductive \emph{B} tactics, not 
available in current \emph{B} implementations.
\item \emph{GSL} definition -- either through a shallow embedding (an approach similar to the
one presented in \cite{bod:1b}, but in \emph{BiCoq}) or through a deep embedding (with 
higher-order and first-order refinement definitions, and proof of equivalence).
\end{itemize}

We would like to emphasise the simplicity and the efficiency of the deep embedding approach, 
when having both validation and implementation objectives. In a relatively short amount of 
time, it was possible to describe the \emph{B} logic, to check its standard results, and to 
implement a proved prover for this logic.

\subsubsection*{\small Acknowledgements.}\label{sc_ack}\small We thank Pr. Hardin for 
reviewing earlier versions of this paper.

%----------------------------------------------------------------------------------------------
% Bibliography
%----------------------------------------------------------------------------------------------

%\begin{thebibliography}{5}
%\end{thebibliography}

\bibliographystyle{splncs}
\bibliography{BiCoq}

%----------------------------------------------------------------------------------------------
% End of the document
%----------------------------------------------------------------------------------------------

\end{document}